\documentclass{ws-procs9x6}

\begin{document}

\title{Low mass stellar evolution with WIMP capture and annihilation}

\author{\underline{P.~C.~Scott}$^*$ and J.~Edsj\"o}
\address{Cosmology, Particle Astrophysics and String Theory, Physics, Stockholm University \&\\
High Energy Astrophysics and Cosmology Centre (HEAC),\\AlbaNova Univeristy Centre, SE-106 91 Stockholm, Sweden\\
$^*$Presenting author; pat@physto.se}

\author{M.~Fairbairn}
\address{PH-TH, CERN, Geneva, Switzerland \& King's College London, WC2R 2LS, UK}

\begin{abstract}
Recent work has indicated that WIMP annihilation in stellar cores has the potential to contribute significantly to a star's total energy production.  We report on progress in simulating the effects of WIMP capture and annihilation upon stellar structure and evolution near supermassive black holes, using the new DarkStars code.  Preliminary results indicate that low-mass stars are the most influenced by WIMP annihilation, which could have consequences for upcoming observational programs.
\end{abstract}


\bodymatter

\section*{}
Our current understanding of cosmology is that over 20\% of the mass-energy in the universe is in the form of `dark matter' \cite{Bergstrom00, Bertone05, Spergel07}, the composition of which remains unknown.  One promising group of dark matter candidates are weakly interacting massive particles (WIMPs), attractive because the masses and couplings associated with the weak scale naturally lead to a relic abundance of dark matter consistent with present-day observations.

WIMPs should posses a small but non-zero weak scattering cross-section with standard model particles.  This means that they could scatter off atomic nuclei in stars, become gravitationally captured and eventually congregate in stellar cores.  WIMP accretion by stars has been studied extensively \cite{Press85, Griest87, Gould87a, Gould87b}, typically with a view to observing neutrinos produced by WIMP self-annihilation in the core of the Sun.  Others have considered the influence of larger concentrations of WIMPs upon the structure of the stars themselves, focusing on conductive energy transport \cite{Spergel85, BouquetSalati89b, Salati90, Dearborn90, Bottino02} and energy production by annihilations \cite{SalatiSilk89, Moskalenko07, Bertone07}.  

We provide preliminary results from simulations of the influence of WIMP capture and annihilation upon main sequence stars, using the evolutionary code DarkStars.  We have also presented results obtained with a simpler static stellar structure code \cite{Fairbairn07}, using some results presented herein for comparison.  DarkStars is built upon the stellar evolution package EZ \cite{Paxton04}, derived from Eggleton's \textsc{stars} code \cite{Eggleton71, Eggleton72, Pols95}, and a generalised version of the capture routines in DarkSUSY \cite{darksusy}, which are based upon the capture expressions of Gould \cite{Gould87b}.  The code includes a detailed treatment of conductive energy transport by WIMPs, using the expressions of Gould \& Raffelt \cite{GouldRaffelt90a} with correction factors derived from their accompanying numerical solutions to the Boltzmann equation \cite{GouldRaffelt90b}.  The WIMP radial distribution is allowed to deviate from a strictly isothermal Gaussian in a manner consistent with the treatment of the conductive energy transport.  DarkStars, its theoretical underpinnings and application to a range of different stars will be described in full in a coming publication \cite{future}.

We focus on conditions obtained near supermassive black holes, where the highest ambient dark matter densities are expected to be found \cite{Merritt04, Merritt05}.  We use the maximal values of both the spin-dependent \cite{Angle07} ($10^{-38}$\,cm$^2$) and spin-independent \cite{Desai04} ($10^{-44}$\,cm$^2$) WIMP-nucleon cross-sections currently allowed by direct and indirect detection experiments, and assume a 100\,GeV WIMP mass.  The annihilation cross-section is set to $3\times10^{-26}$\,cm$^3$\,s$^{-1}$, as demanded by relic abundance considerations \cite{Spergel07}.  For simplicity we currently work with dark matter halo parameters for the Sun, assuming a Gaussian WIMP velocity distribution of width 270\,km\,s$^{-1}$ and a stellar proper motion of 220\,km\,s$^{-1}$ relative to the halo.  We assume a metallicity of $Z=0.02$, which is in the vicinity of the Sun's \cite{AspRev}.

\begin{figure}
\begin{center}
\psfig{file=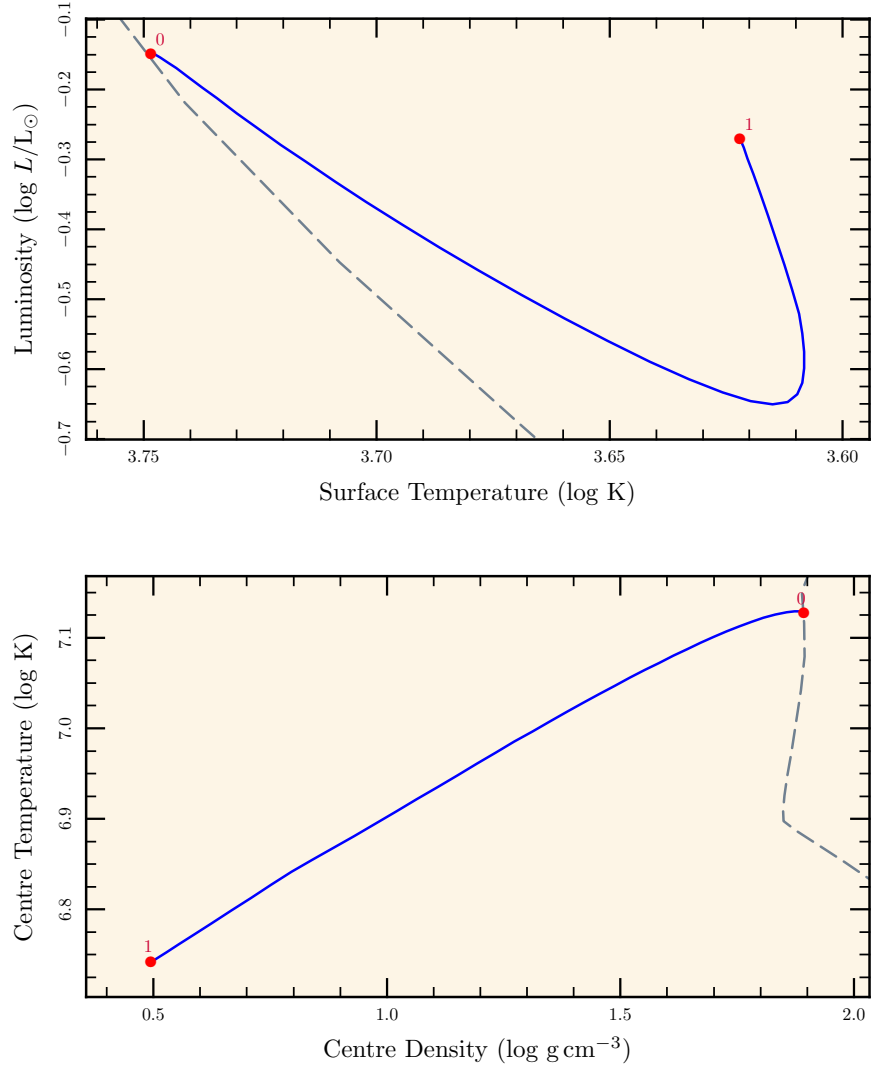,width=\textwidth}
\end{center}
\caption{The path followed in the HR (top) and central equation of state (bottom) diagrams by a 1.0\,$M_\odot$ star, evolved for 100\,Myr in a 10$^{10}$\,GeV\,cm$^{-3}$ WIMP halo.  The dashed lines indicate the zero age main sequence, which defines the boundary for hydrogen-burning in the lower plot (to the left of the line, hydrogen fusion cannot occur).  The red points indicate the starting (0) and final (1) positions.}
\label{fig1}
\end{figure}

\begin{figure}
\begin{center}
\psfig{file=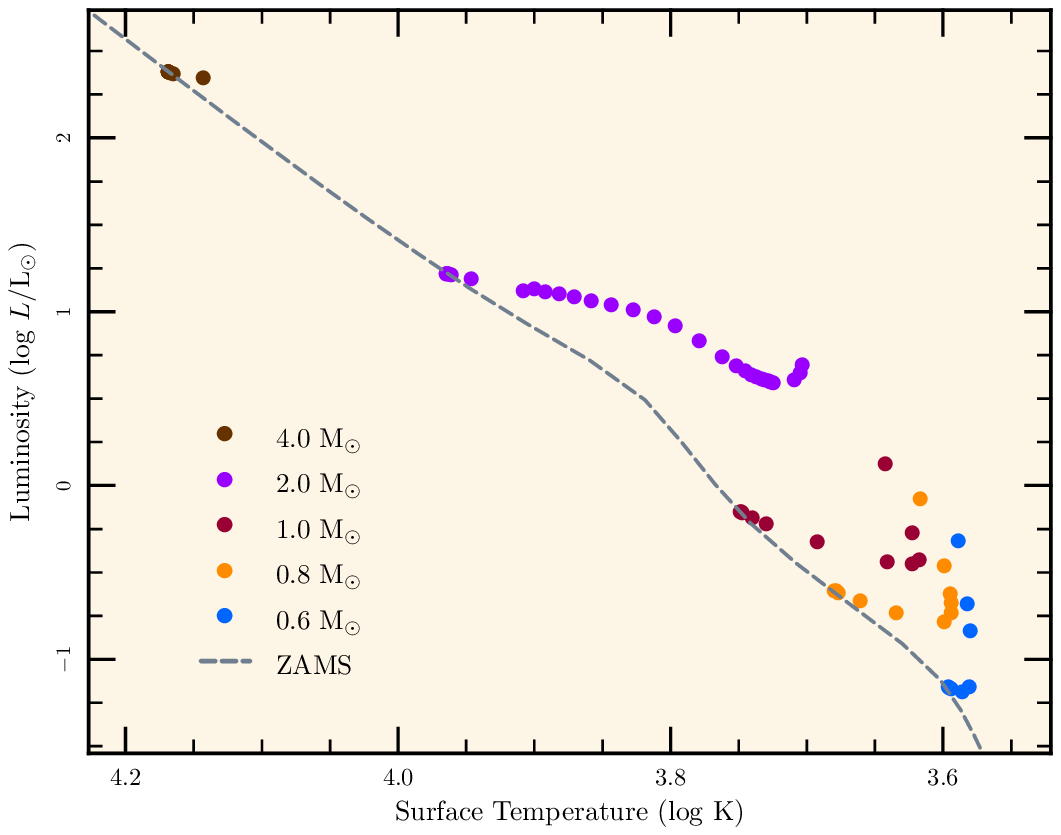,width=\textwidth}
\end{center}
\caption{The positions in the HR diagram at which main sequence stars stabilise when embedded in WIMP halos of 10$^7$ -- $5\times10^{10}$\,GeV\,cm$^{-3}$.  Points closest to the main sequence are those computed with 10$^7$\,GeV\,cm$^{-3}$ halos.  For every mass shown, points are given over the entire range of WIMP densities, but not necessarily with the same resolution (e.g.~more points were computed in the 2.0\,$M_\odot$ case).}
\label{fig2}
\end{figure}

Given a large, constant ambient WIMP density, capture and annihilation quickly equilibrate inside stars, such that annihilation provides a roughly constant source of additional energy.  Beacuse the WIMPs congregate very close to the stellar centre, this energy is produced in a much smaller region than that from nuclear burning.  The initial, concentrated, rapid injection of energy raises the luminosity gradient in the core of a star, steepening the temperature gradient and prompting the creation of a central convection zone (or the rapid expansion of the existing one).  The increased efficiency of energy transport causes the core to cool and expand, increasing the stellar radius and decreasing surface temperature and central density.  The top panel of Fig.~\ref{fig1} shows the evolution of a 1.0\,$M_\odot$ WIMP burner in the HR diagram as it adjusts to the presence of the extra energy in its core from WIMP annihilations.  The lower panel shows the corresponding changes in the central equation of state (temperature and density) during this process.  Nuclear burning eventually switches off as the central temperature and density become too low to support it, leaving WIMPs to power the star alone.  

The surface cooling brought about by the star's expansion allows H$^-$ ions to survive to increasing depths, with the resulting opacity increase causing the surface convection zone to expand.  If enough WIMPs are present, the expanding surface and core convection zones will meet.  This results in a fully convective star with a total luminosity which rises once again, as the further enhancement in energy transport allows the energy reaching the surface to outstrip that powering the overall expansion.  Figure \ref{fig2} shows the location in the HR diagram of main sequence stars of various masses in differing ambient dark matter densities from 10$^7$ -- $5\times10^{10}$\,GeV\,cm$^{-3}$, after having completely adjusted to the effects of the WIMPs.  The change in direction of the `tracks' approximately corresponds to the point at which nuclear burning turns off entirely, the star becomes fully convective and its luminosity undergoes some increase after the initial decrease.  

The effects of WIMPs upon main sequence stars are most pronounced at low stellar masses, simply because the energy from nuclear burning scales as roughly the third or fourth power of $M$, whereas the WIMP capture rate is almost linear in $M$.  Since WIMPs are in principle an eternal source of energy, WIMP burners will shine and occupy the same position in the HR diagram indefinitely.  (After being evolved for a further 30\,Gyr beyond what we show here, the star in Fig.~\ref{fig1} was virtually indistinguishable from itself at age 100\,Myr.)  This suggests that main sequence WIMP burners could be found by  examining regions where stars cannot have formed recently, looking for populations where lower mass stars appear oddly younger than higher mass ones.  Whilst the stars we describe are far too cool to solve the `paradox of youth' reported at the centres of M31 and our own galaxy \cite{Ghez03,Demarque07}, some of the explanations for this paradox demand the presence of a fainter, as-yet unobserved population of lower-mass stars.  If main sequence WIMP burners exist anywhere, such stars would be prime examples.  Since upcoming observations of the galactic centre will soon reach the sensitivity required to detect such a population, discovery of WIMP burners might be just around the corner.

\bibliography{DMbiblio}

\end{document}